\documentclass{easychair}

\usepackage{doc}

\usepackage{booktabs}
\usepackage{fancyvrb}

\usepackage[overlay,absolute]{textpos}
\newcommand\PlaceText[3]{%
\begin{textblock*}{10in}(#1,#2)  
#3
\end{textblock*}
}%
\textblockorigin{-5mm}{0mm}   

\title{Scamming the Scammers: Using ChatGPT \\ to Reply Mails for Wasting Time and Resources}

\author{
Enrico Cambiaso
\and
    Luca Caviglione
}

\institute{
  National Research Council of Italy, Genova, Italy\\
  \email{\{name.surname\}@cnr.it}
}

\authorrunning{Cambiaso and Caviglione}

\titlerunning{Scamming the Scammers}

\begin{document}

\PlaceText{-17mm}{10mm}{\centering{This paper has been submitted for publication in\\ ITASEC23 - The Italian Conference on Cybersecurity, May 3rd - 5th, 2023, Bari, Italy.}}
\maketitle

\begin{abstract}
The use of Artificial Intelligence (AI) to support cybersecurity operations is now a consolidated practice, e.g., to detect malicious code or configure traffic filtering policies. The recent surge of AI, generative techniques and frameworks with efficient natural language processing capabilities dramatically magnifies the number of possible applications aimed at increasing the security of the Internet. Specifically, the ability of ChatGPT to produce textual contents while mimicking realistic human interactions can be used to mitigate the plague of emails containing scams. Therefore, this paper investigates the use of AI to engage scammers in automatized and pointless communications, with the goal of wasting both their time and resources. Preliminary results showcase that ChatGPT is able to decoy scammers, thus confirming that AI is an effective tool to counteract threats delivered via mail. In addition, we highlight the multitude of implications and open research questions to be addressed in the perspective of the ubiquitous adoption of AI.  
\end{abstract}

\section{Introduction} \label{sec:introduction}

The use of mails to perform scams, drop attack payloads, deliver malicious URLs, and distribute unwanted spam messages has been a prime vector used by attackers since the early days of the Internet. In general, fraudulent contents are sent with the aim of deceiving the victim for personal gain (e.g., to receive moneys) or to force some behaviour (e.g., to install an executable). With the increasing diffusion of the Internet, the impact of threats delivered via mail is now very relevant, both considering the economic losses for the victims and the effort dedicated to detect harmful messages or attachments~\cite{jones2019email}. As today, the overall fraction of mails supporting frauds and criminal activities is up to the $90\%$ of the total exchanged volume and this trend is expected to grow in the near future~\cite{burke2021prepare, aiphish, kovalluri2018lstm}. Therefore, mitigating the impact of malicious and unwanted mails is a crucial activity, not only limited to human aspects but also to prevent waste of resources (e.g., bandwidth and storage of mail servers). Among the various techniques proposed to counteract the plague of frauds or attacks delivered through mails, a vast corpus of works dealing with the use of Artificial Intelligence (AI) has  emerged~\cite{spamsurvey}. For instance, AI can be used to detect malicious mails, create filters, or even generate automatic replies. In this vein, our work aims at evaluating whether some form of AI can be used to interact with scammers and attract them in unproductive conversations.

Specifically, engaging scammers requires to generate suitable replies. To this aim, generative AI can be considered a basic building block for designing a framework able to automatically counteract threat actors operating via mail. In fact, generative techniques are capable of exploiting a knowledge set to generate novel contents~\cite{gozalo2023chatgpt}. For instance, models like Stable Diffusion or Dall{$\cdot$}E 2 can produce images starting from text~\cite{borji2022generated}, whereas other tools can be used to create multimedia objects, such as music or videos~\cite{muller2022genaichi}. With the goal of generating convincing replies to scam messages, ChatGPT (\url{https://chat.openai.com}) seems one of the most promising and interesting methods. In essence, it implements a Natural Language Processing (NLP) generative algorithm developed by OpenAI to mimic realistic interactions during general purpose conversations~\cite{gozalo2023chatgpt}. Launched in November 2022, ChatGPT quickly gained popularity, reaching $1$ million of total users in just $5$ days\footnote{To roughly quantify the disruptive potential of ChatGPT, its diffusion can be compared with other Internet-wide services. Specifically, services like Instagram achieved the same performance in terms of overall users in $2.5$ months, whereas Netflix took $3.5$ years. For a detailed report, see: \url{https://www.statista.com/chart/29174/time-to-one-million-users/} (Last accessed on February 10, 2023)}. In the wake of its popularity, ChatGPT has been investigated both by the industry and academia to create a wide range of contents. For instance, it has been used to write convincing scientific papers~\cite{zaremba2023chatgpt}, to support medical patients by providing easy to understand reports~\cite{jeblick2022chatgpt}, to act as a network honeypot~\cite{mckee2023chatbots}, as well as for specific tasks such as the generation of code snippets or the early detection of security vulnerabilities~\cite{aljanabi2023chatgpt}.

Owing to its flexibility, this paper aims at evaluating the use of ChatGPT as a synecdoche of generative techniques to counteract the plague of mail scams. Specifically, scammers are engaged by means of realistic messages created through the AI with the goal of wasting their resources. Even if the limit of our investigation relies on the small number of considered attacks, the main goals of the paper are understanding the feasibility of the approach and outlining the perspective issues and research gaps to be addressed in the near future. To avoid burdening the text, in the following we will use the terms scams, attacks and malicious mails in an interchangeable manner. However, when doubts may arise, we will specify the type of threat, e.g., spam or phishing. 

Summing up, the contributions of this work are: \textit{i}) understanding the feasibility of using ChatGPT as a ``security'' tool to counteract malicious mail messages, \textit{ii}) providing a preliminary quantitative assessment of the effectiveness of the AI-based approach, and \textit{iii}) shaping the main research questions and engineering challenges to be addressed in the perspective of using generative methods to counteract mail-based scams. 

The rest of the paper is structured as follows. Section~\ref{sec:relatedwork} presents the previous works dealing with the adoption of AI to counteract various types of unsolicited mails. Section~\ref{sec:methodology} discusses the framework and methodology used to prove the effectiveness of ChatGPT to generate coherent answers, while Section~\ref{sec:results} showcases numerical results obtained via preliminary tests. Section~\ref{sec:questions} deals with some research questions that should be addressed and, finally, Section~\ref{sec:conclusions} concludes the paper and outlines possible future works.

\section{Related Work} \label{sec:relatedwork}

Mail messages are regularly abused to deliver a wide range of threats and they are one of the preferred vector to deploy ransomware attacks~\cite{mailransom}. Besides, the majority of messages are devoted to support phishing campaigns or spam communications~\cite{aiphish}. Indeed, mails are the main mechanism for implementing different and sophisticated fraud schemes to extort money~\cite{mailfraud}.
In more detail, scam attempts tend to cluster into several, recurrent categories, such as messages threatening the victim or asking for charity~\cite{tuvser2022trends}. However, the most popular and effective scam messages refer to large winnings notifications~\cite{chiluwa2019congratulations}. As a consequence of the massive diffusion of mail communications, the design and deployment of efficient protection mechanisms have been prime research topics for several decades and still pose many open research challenges, especially due to adversaries continuously evolving and adapting their offensive strategies. 

As regards the mitigation of unwanted mails and scam messages, the literature proposes several approaches. For instance, \cite{roman2005protection,roman2006anti} showcase a challenge-response scheme that the sender has to complete before contacting the recipient, i.e., to be whitelisted and avoid further checks. Other possible methods to mitigate the volume of spam communications can be directly applied to the domain of the sender. In more detail, the Sender Policy Framework and the DomainKeys Identified Mail can be used to prevent spammers from sending messages through a well-defined domain also by means of spoofed identities~\cite{spamsurvey}. Unsolicited and malicious contents can also be counteracted at a protocol-level. In this case, \cite{park2022advanced} proposes an extension to the SMTP to automatically check whether the domain of the sender corresponds to a valid DNS entry. The impact of fraud mails can also be assessed by considering the content of the message. As an example, \cite{kim2019scam} identifies scam communications through text analysis, i.e., inappropriate statements are identified.

More recently, techniques to reduce the impact of malicious mails are increasingly exploiting AI or machine-learning-capable approaches. In this regard, \cite{yaseen2021spam} considers deep learning algorithms to classify spam messages through word embedding techniques, while \cite{salem2010awareness} proposes a real-time detection system for the identification of phishing attacks. To the best of our knowledge, previous techniques for mitigating the impact of scam attempts via mail do not consider the use of generative AI-based schemes to engage scammers. The only notable exception is \cite{kovalluri2018lstm}, although it adopts a long short-term memory approach to generate basic questions and consume the time of the attacker. Concerning AI techniques to implement spam/scam countermeasures, they have been primarily used to automatically inspect various parts of a message in order to detect spam or phishing mails, i.e., for classification purposes. Specifically, the AI can be used to check the headers, the SMTP envelope, or different portions of SMTP data~\cite{aiphish,spamsurvey}.

Employing AI to generate mail contents or to face some security issues has already been partially investigated, even with scopes different from those addressed in this paper. In more detail, \cite{karanjai2022targeted} exploits natural language models (i.e., GPT-2 and GPT-3) to generate email phishing messages to conduct tests. Besides, \cite{mckee2023chatbots} showcases how ChatGPT can be adopted to simulate Linux/Windows terminals within a honeypot.

\section{Methodology} \label{sec:methodology}

To evaluate the feasibility of taking advantage of AI to interact with scammers, we prepared a simple testbed. First, we selected a mail account with a realistic domain (i.e., {\tt @cnr.it}), which has been publicly available on the Internet for years. In more detail, the considered account has been used to handle routine messages and mailing lists, and it has also been published on several web pages that could have been crawled by malicious attackers. To operate the mail account we used the Microsoft Office365 platform, which includes an anti-spam filter.

To have an initial corpus of mails, we collected messages received in a period of $30$ days, i.e., from 12th of November 2022 to 12th of December 2022. The overall experimentation lasted $60$ days, i.e., from 12th of November 2022 to 11th of January 2023. Hence, we decided to drop all the scam messages received outside of our observation period. Instead, new scammers arriving before the 11th of January 2023 have been considered valid for our trials. To identify scammers, we used the following approach. Mails flagged as malicious by the Office365 platform have been manually inspected to evaluate their inclusion in our test set. For the sake of our investigation, we did not consider mails containing phishing attempts or those mimicking popular services or HTML pages requiring to directly follow a link \cite{spamsurvey}. Instead, we only considered plain-text messages asking for a direct interaction, i.e., a reply. 

To generate replies we used ChatGPT. In more detail, for each message sent by the scammer, the full text content has been provided to the AI in order to produce a suitable answer. Unfortunately, at the time of our experiments, ChatGPT does not allow to perform tweaks or to alter its normal behavior, i.e., it must be considered as a black-box solution. As a consequence, directly feeding the AI with scam messages led to a warning without providing a suitable answer. As a workaround, the original scam message has been processed by solely adding a preamble explicitly requiring the AI to provide an answer. Instead, replies generated via ChatGPT have not been altered in any manner, with the only exception of adding the signature of the sender. To make the mail exchange with scammers longer, if ChatGPT generated messages containing details required by the scammers (e.g., bank account information, postal addresses, or telephone numbers), we again tweaked the preamble to instruct the AI to not provide any personal detail. 

For reducing the chances that the scammer could spot the ``unmanned'' nature of the replies, we mimicked the presence of a human endpoint by delaying the various answers. Then, in our trials we provided replies by randomly waiting for a period, which ranges from minutes to weeks.

Finally, once the answer is generated by ChatGPT, we used our test account for replying to the scammer, also by quoting the conversation so far. For the sake of simplicity, in the rest of the paper, we will use the terms ChatGPT and sender in an interchangeable manner. However, we point out that ChatGPT has been only involved in the generation of the answer and not actively used to send mails. 

\section{Preliminary Results} \label{sec:results}

\begin{table}[tb]
    \small
    \centering
    \begin{tabular}{c|c|c|c|c|c|c}
         \toprule
         {\textbf{Scammer}} & {\textbf{SMTP Status}} & {\textbf{Thread}} & \multicolumn{2}{c|}{\textbf{Scammer Msg. Len.}} & \multicolumn{2}{c}{\textbf{ChatGPT Msg. Len}} \\
         ID & Code & No. Mails & Avg. Chars & Avg. Sent. & Avg. Chars & Avg. Sent. \\
         \midrule
         1 & Failed (\texttt{5.2.1}) & 2 & 331 & 2 & 333 & 3 \\ 
         2 & Failed (\texttt{5.2.1}) & 2 & 261 & 1 & 323 & 4 \\ 
         3 & Failed (\texttt{5.2.1}) & 2 & 291 & 3 & 362 & 4 \\ 
         4 & - & 12 & 1,487 & 13 & 536 & 6 \\ 
         5 & - & 2 & 4,572 & 48 & 319 & 5 \\ 
         6 & - & 10 & 987 & 6 & 526 & 6 \\ 
         7 & - & 2 & 11,094 & 108 & 487 & 5 \\ 
         8 & - & 14 & 1,382 & 12 & 473 & 5 \\ 
         9 & - & 2 & 120 & 1 & 277 & 5 \\ 
         10 & - & 18 & 474 & 7 & 432 & 6 \\ 
         11 & - & 2 & 207 & 7 & 292 & 7 \\ 
         \bottomrule
    \end{tabular}
    \caption{Overall volume statistics of the threads between scammers and ChatGPT. Scammers have been sorted according to the date of the first received mail message.}
    \label{tab:volume}
\end{table}

Table~\ref{tab:volume} contains volume statistics of the various messages exchanged between a specific scammer and the ChatGPT instance. Concerning Scammer 1, 2, 3, and 9, the mail thread stopped after a single reply from ChatGPT. As shown, for Scammer 1, 2, and 3,  the sent messages have not been received, since we obtained an error, i.e., \texttt{SMTP status 5.2.1}. However, this behavior is something that could be expected and should not be considered a flaw in the ``credibility'' of the message generated through the AI. Moreover, the original messages from Scammer 1, 2, 3, and 9, contained a .pdf attachment, thus they were probably intended as a one-shot communication to drop a payload on the host of the victim, e.g., a keylogger, or to support a web phishing campaign~\cite{spamsurvey}.
A similar explanation holds for Scammer 11. In this case, no errors were generated by the mail provider, but the original communication contained a link to a website (i.e., \texttt{http://KAV[SANITIZED].NET}) and a set of credentials. For the case of Scammer 5 and 7, we did not receive any answer as well, but the replies have been successfully delivered. Most probably, the malicious actor stopped surveying the address in the meantime, for instance to avoid detection or because he/she has been neutralized. 

With Scammer 4, ChatGPT exchanged 6 different mails, leading to an overall thread of 12 messages, while with Scammer 6 the AI has been used to lock the malicious actor in a thread of 10 mails. Concerning the effectiveness of ChatGPT to generate realistic messages, it is worth considering the conversation with Scammer 8, which happened across Christmas holidays. After exchanging 12 mails, the attacker wrote a message for making wishes. Then, he/she asked for a telephone number to switch the conversation from mail to voice\footnote{Actually, the scammer further replied, but we did not consider the message since it was outside our window of observation.}. See Table~\ref{tab:sample} for details on the textual form/content of the various mail messages. It is worth underlining that organizing a phone meeting has been proposed by ChatGPT in its reply. This is surely a limit of using a tool trained over a general corpus of text, rather than on a dataset properly suited for counteracting malicious mails. Lastly, Scammer 10 is the one exhibiting in the highest number of replies, i.e., we exchanged 18 mails in total. It is worth mentioning that after some mails generated via ChatGPT, the malicious actor decided to ask money by sending pointers to a Western Union account. As today, the scammer is still hanged waiting for a reply or an acknowledgement of the wire transfer. Similarly, Scammer 4 is waiting for a feedback about a money transfer through SilverBank.

\begin{table}[tb]
    \centering
    \begin{tabular}{c|c}
    \toprule
    \textbf{ChatGPT-generated Message} & \textbf{Scammer Reply}\\
    \midrule
      \begin{minipage}[t]{0.56\textwidth}
      \begin{Verbatim}[fontsize=\footnotesize]
Dear Mr. [OMITTED],

Thank you for reaching out to me regarding the cash 
deposit and properties left behind by my late relative. 
I am interested in proceeding with making a claim on 
these assets and would appreciate any assistance you 
can provide.

I understand that you need my personal information 
in order to facilitate the process. Please note that 
I am unable to disclose my full name, address, or any 
other personal information via email. However, I am 
happy to speak with you by phone or schedule a 
meeting in person to discuss this further.

Please let me know how you would like to proceed.

Best regards,
[OMITTED]

      \end{Verbatim}
      \end{minipage}   &
      \begin{minipage}[t]{0.39\textwidth}
      \begin{Verbatim}[fontsize=\footnotesize]
Dear [OMITTED],

Compliments of the Season. 
I am happy hearing from you as regards 
to this process. Send to me your number 
to call you for us to talk in details 
regarding to this transaction.

thanks,
Kar[OMITTED].
      \end{Verbatim}
      \end{minipage} \\
      \bottomrule
    \end{tabular}
    \caption{Mail exchange between ChatGPT and Scammer 8 across the holiday season. (Notice: minor formatting has been done to better fit the text in the table).}
    \label{tab:sample}
\end{table}

As shown in Table~\ref{tab:volume}, each thread could also be characterized by using simple metrics. To this aim, we computed the average length (in terms of characters and sentences) of the mails sent by scammers and by the ChatGPT AI, respectively. In more detail, the average length of messages sent by Scammer 5 and 7 is $4,572$ and $11,094$, respectively. However, the threads stopped after only two messages, probably since the malicious actors expected a different type of answers despite their length (the highest in our trials). Therefore, answer messages generated by ChatGPT should be somewhat also tweaked to further engage the scammer or not raise suspects. By analysing the content of both the emails received, the malicious senders suggest to invest moneys in specific stocks, thus asking for precise witnesses that the victim has been successfully decoyed. For the other threads, we did not found any relevant correlation between the length of the messages and the behavior of the scammer (specifically, in terms of the number of exchanged mails). 

\begin{figure}[tb]
    \centering
    \includegraphics[width=0.74\textwidth]{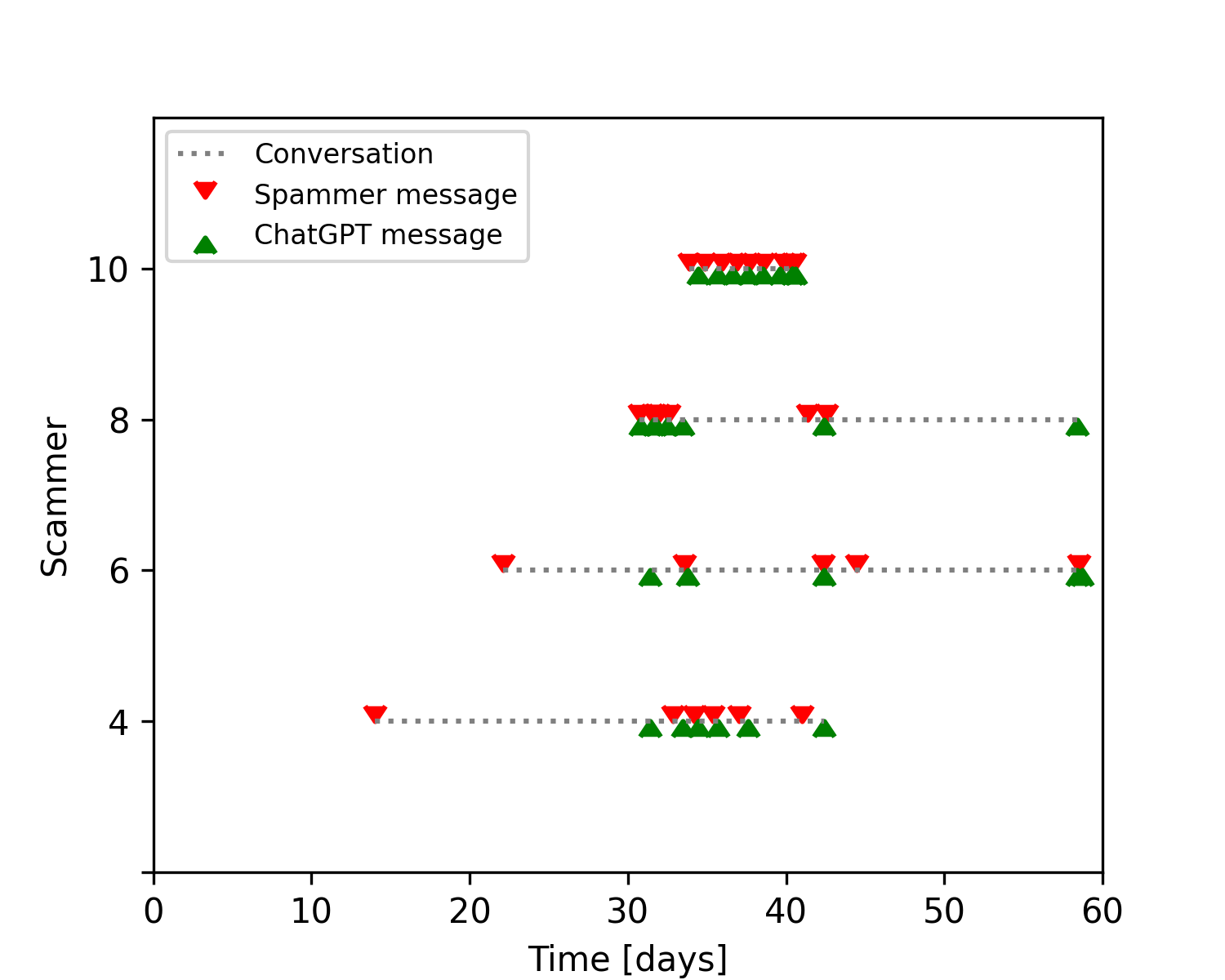}
    \caption{Overview of the conversations with scammers who ``play the game'' during the observation period.}
    \label{fig:graph}
\end{figure}

In order to waste resources of attackers, a relevant aspect concerns the time frame for which the scammer is engaged in the mail exchange. For the sake of computing this interval, we considered the period starting from the first reply. 
Figure~\ref{fig:graph} depicts the evolution of the various mail threads handled via ChatGPT during our observation window. In general, messages prepared with ChatGPT engaged spammers for $\sim$$18$ days, on average. Specifically, the exchange with Scammer 10 was the shorter and lasted $\sim$$6$ days. In this case, the mail flow was quite tight (i.e., $2$ mails per day) to pursue the malicious goal of providing  coordinates for the wired transfer as soon as possible. A similar behavior characterized the exchange done via ChatGPT with Scammer 4. To evaluate the impact of the initial response time, in this case we delayed the reply to the first message of $17$ days. Instead, for Scammer 8 and 6 the conversations were longer, i.e., $\sim$$27$ days. With Scammer 8 we exchanged mails in a ``bursty'' manner interleaved with two stops period of $8$ and $16$ days, respectively. For Scammer 6 the exchange was more regular with two bursts of messages sent after a stop period of $9$ and $14$ days. 

\section{Open Research Questions} \label{sec:questions}
 
As hinted, the use of some form of AI to mitigate the various types of threats targeting mails (e.g., fraud, phishing, spam, or the drop of malicious payloads) has become a vivid research topic \cite{spamsurvey}. Roughly speaking, the most recent efforts seem to cluster around two major topics. The first aims at advancing in filters used to classify and detect mails, especially with the aim of preventing attacks or feeding a cybersecurity framework, for instance to automatically quarantine attachments or bounce messages (see \cite{aiphish} for the case of phishing). The second aims at understanding the potential of AI when deployed to support or substitute humans. As an example, phishing messages can be automatically generated by using AI to train users against social engineering attacks \cite{aigenmail}. Despite the goal, a relevant shared portion of research requires to fully assess the multifaceted set of implications of mixing machine learning with security frameworks or countermeasures for inter-personal communications \cite{aiphish, aigenmail, spamsurvey}.

In this perspective, the use of ChatGPT opens different research questions often requiring to deal with the multifaceted flavor of AI and its rapidly-evolving pace. Thus, successfully incorporating AI in production-quality security frameworks requires to consider human and ethical aspects, computational optimizations and explainability constraints \cite{aicite}. Specifically, the main open research questions that we identified when conducting our experimentation are:

\begin{itemize}
    \item \textbf{Specialized and as-a-Service Implementations}: in general, it is hard to forecast a one-size-fits-all mechanism able to face with the various hazards delivered via mail, e.g., whale phishing or drop of malicious payloads. Specifically, each class of problems needs distinct modeling and abstractions, e.g., the allotted vocabulary (see \cite{mailfaq} for specific traits of mails contributing to automatically generated FAQs). Unfortunately, the deployment of a framework for answering to an overwhelming amount of scam messages in an automatic manner could not be feasible for many small/medium-sized entities. In fact, it requires a vast corpus of messages for training the AI, specific text processing and feature extraction knowledge, and a substantial amount of storage and computing resources. Thus, industrial and academics should work towards implementations offered as-a-Service to take advantage of scale factors, especially to have enough data to train and tweak models. 

    \item \textbf{Modeling the Human Behavior}: even if the text provided by ChatGPT (or other forms of AI) could appear as sound and valid, the scammer could detect the lack of a human counterpart due to patterns in text, absence (or presence) of grammatical errors, or too fast replies. In this vein, inspecting the received mail messages could be used by attackers to perform reconnaissance and fingerprint AI endpoints \cite{reconcav}. Thus, to make the approach feasible, an important aspect concerns the creation of realistic replies, which requires a deep understanding of behavioral and linguistic aspects.
      
    \item \textbf{Privacy and Forensics}: automatic and AI-driven mail answering requires to gather a relevant amount of real messages in order to generate suitable replies. This could clash with privacy requirements and regulations such as the General Data Protection Regulation increasingly pushing to the minimum the needed information \cite{gdpr}. Moreover, increased volumes of AI-generated replies could lead to difficulties in performing forensics investigations or tracing scammers across multiple services producing messages \cite{forensics}. Thus, suitable tradeoffs between privacy, rights of users, and performances should be searched for. 

    \item \textbf{Avoid Unwanted Traffic}: the massive deployment of AI-based countermeasures is expected to exacerbate the automation of many security processes, while minimizing the presence of a human in the loop. At the same time, it is unlikely that threat actors will not take advantage of AI or machine-learning-capable tools to generate messages or handle responses. Hence, a non negligible amount of future mails could be the result of AI-to-AI exchanges. As a consequence, part of the ongoing research should also consider suitable techniques to mitigate the plague of unwanted traffic accounting for resource wastes and economical losses \cite{unwanted}.

    \item \textbf{Ethical Implications}: interacting with humans and handling people-centric data and communications rise several ethical concerns. First, the idea of using AI to ``scam the scammers'' is somewhat intrinsically a fraud, since it goes beyond the classification of messages or the detection of malicious contents. Second, a plausible corpus of mails could contribute to spread untrue statements or exacerbate issues in discriminating contents created by humans from those generated by machines (see, e.g., the case of using ChatGPT for online exams or general education duties   \cite{susnjak2022chatgpt,zhai2022chatgpt}). 
\end{itemize}

Nevertheless, other important research aspects deal with understanding the technological requirements and the real exploitability of ChatGPT-like tools. In fact, despite being preliminary, our current experimentation did not take advantage of ad-hoc tools. Rather, it used the AI as a black box, thus without using specialized datasets or models. 
At the same time, both the research and industrial communities should take in high regards the open points outlining the shape of AI-based generative mechanisms. For instance, Internet services should be able to block messages generated by the AI to not spread unrealistic messages (e.g., as it happens for StackOverflow posts \cite{weisz2023toward}) or to avoid that an attacker can steal the model and use it for weaponizing his/her attack campaigns. 

\section{Conclusions and Future Works} \label{sec:conclusions}

In this paper we presented the use of ChatGPT to generate email messages to engage scammers and waste their resources. Results indicated that AI can be a valuable and effective tool, as we were able to exchange up to $18$ mails with a single scammer, or to trick attackers for up to $27$ days. Even if our experimentation was limited, it allowed to highlight the multitude of research and ethical questions that the use of a framework like ChatGPT rises. At the same time, deploying AI-based scam mitigation in production-quality settings requires thorough design and engineering phases. This paper should then be considered a sort of ``manifesto'' of the multifaceted and complex alchemy arising from the mix of personal mails messages and generative AI. 

Future works aim at extending the scope of the experimentation both in terms of volume of mails considered (e.g., threads and senders) and the impact of the semantic/contents of the used text. Another relevant part of our research is devoted to improve our testbed to reduce at the minimum the need of human support, e.g., by integrating the mail management framework with ChatGPT/AI.

\section*{Acknowledgment}

This work has been partially supported by SERICS - Security and Rights in CyberSpace (\url{https://serics.eu}), within the Piano Nazionale di Ripresa e Resilienza, funded by the NextGenerationEU framework (No. 341 - 15/03/2022). 

\label{sect:bib}
\bibliographystyle{plain}
\bibliography{biblio}

\end{document}